\documentclass[reprint]{JASA}

\begin{document}
\title[Kirchhoff Integral Approximation]{A series approximation to the Kirchhoff integral for Gaussian and exponential roughness covariance functions}
\thanks{Published in the Journal of the Acoustical Society of America. The published version of this preprint can be found at \url{https://doi.org/10.1121/10.0005282}}
\author{Derek R. Olson}

\affiliation{Oceanography Department, Naval Postgraduate School, Monterey, CA 93943}
\date{\today}
\begin{abstract}
The Kirchhoff integral is a fundamental integral in scattering theory, appearing in both the Kirchhoff approximation, as well as the small slope approximation. In this work, a functional Taylor series approximation to the-eps-converted-to.pdf Kirchhoff integral is presented, under the condition that the roughness covariance function follows either an exponential or Gaussian form--in both the one-dimensional and two-dimensional cases. Previous approximations to the Kirchhoff integral [Gragg et al. J. Acoust. Soc. Am. 2001, Drumheller and Gragg J. Acoust. Soc. Am., 2001] assumed that the outer scale of the roughness was very large compared to the wavelength, whereas the proposed method can treat arbitrary outer scales. Assuming an infinite outer scale implies that the root mean square (rms) roughness is infinite. The proposed method can efficiently treat surfaces with finite outer scale, and therefore finite rms height. This series is shown to converge independently of roughness or acoustic parameters, and converges to within roundoff error with a reasonable number of terms for a wide variety of dimensionless roughness parameters. The series converges quickly when the dimensionless rms height is small, and slowly when it is large.
\end{abstract}

\maketitle
%\onecolumn
\section{Introduction}
The Kirchhoff integral is extensively used in rough surface scattering theory, appearing in the Kirchhoff approximation for acoustic waves \cite{Thorsos1988,Jackson2007,Ishimaru1978a} and electromagnetic waves \cite{Beckmann1987}, and the small-slope approximation for acoustic waves \cite{Voronovich1985,Voronovich_1994, Yang1994, Gragg2005,Jackson2007,Jackson2020} and electromagnetic waves \cite{Voronovich_1994,Berrouk2014,Afifi2014}. This integral is oscillatory, exists over a semi-infinite domain, and decays slowly. It is thus computationally expensive to evaluate numerically. Fast evaluations of the Kirchhoff integral are needed for performance estimation of active sonar systems \cite{Ainslie2010,Lurton2010}, or inversion of measurements for geoacoustic and roughness parameters using scattering models
\cite{Sternlicht2003a,Steininger2013,Hefner2015}.

Several efforts have been made to approximate this integral. \cite{Gragg2001} and \cite{Drumheller2001}, use a von K{\'a}rm{\'a}n spectral model, but take the limit that the wavenumber cutoff parameter tends to zero, resulting in a pure power law spectrum. Two complementary series approximations are presented in their work, a Taylor series applicable for small roughness and small grazing angles, and a rational fraction approximation that has complementary applicability. In many cases, the domain of applicability of these two series overlaps, allowing accurate calculation of the Kirchhoff integral. This method is useful for cases in which the outer length scale of a rough surface is very large compared to the wavelength. However, the resulting surface statistics have infinite outer length scale, and infinite rms height. Thus modeling scattering from rough surfaces with finite values of these two parameters is at present only possible using brute-force numerical integration. Another series approximation was derived earlier by \cite{Dashen1990} for a pure power law with a spectral exponent of 7/2.

In this work  a functional Taylor series is used to approximate the Kirchhoff integral. This approximation is only directly useful if integer powers of the roughness autocovariance have analytical Fourier transforms (for one-dimensional (1D) roughness), or Fourier-Hankel transforms (for two-dimensional (1D) roughness). Attention is thus restricted to two specific forms of the roughness covariance: the exponential and Gaussian forms. Other forms of the covariance, such as a modified power-law, have this property, but are not commonly used as rough seafloor or terrain models. If the series that is derived here is truncated by retaining the first two terms, then this corresponds to the the reduction of the small-slope approximation to the small roughness perturbation approximation, as noted by several authors, e.g. \cite{Thorsos1995,Voronovich_2002}. That particular simplification is valid for arbitrary covariance functions, but has the same limitations as for the small-roughness perturbation approximation \cite{Thorsos1989}. The series presented here is similar to the small angle series from \cite{Gragg2001} for a pure power-law roughness spectrum. For the series in \cite{Gragg2001}, if the exponential in the Kirchhoff integral is expanded in a power series in $u$, the integration variable, then the resulting series terms are integer powers of the roughness structure function, which is related to the roughness covariance function, since the structure function obeys a power law. Since the power law has no wavenumber cutoff, and no outer scale, the series used in \cite{Gragg2001} converges very slowly for large roughness and steep angles. The series approximation developed in this work is not restricted to small values of the rms height, but is specialized to two classes of covariance functions. It also converges reasonably fast for large rms height and large grazing angles.

In section \ref{sec:definitions}, the basic geometrical and roughness definitions are presented, along with the Kirchhoff integral for both 1D and 2D roughness. The series approximation to the Kirchhoff integral is derived in section~\ref{sec:series} for 1D/2D roughness and exponential/Gaussian covariance functions. The convergence and accuracy of this approximation is studied in Section ~\ref{sec:convergence}, where it is demonstrated that this series converges independently of the roughness or acoustic parameters. It is also shown numerically that a reasonable number of terms is required for a wide variety of dimensionless roughness parameters. Conclusions are given in section~\ref{sec:conclusion}.

\section{Basic Definitions}
\label{sec:definitions}
Here, the basic acoustical quantities that are needed for the Kirchhoff integral are defined. The sound speed in the water is $c_w$, the acoustic frequency is $f$, and the acoustic wavenumber in water is $k_w = 2\pi f/c_w$. The incident and scattered grazing angles are $\theta_i$ and $\theta_s$, respectively. The incident and scattered azimuthal angles are $\phi_i$ and $\phi_s$ respectively. These angles are related to the incident and scattered wave vectors in 3D scattering geometry by, $\mathbf{k}_i = (k_{ix},k_{iy},k_{iz})$, $\mathbf{k}_s = (k_{sx},k_{sy},k_{sz})$, and
\begin{align}
  k_{ix} &= k_w \cos\theta_i \cos\phi_i & k_{sx} &= k_w \cos\theta_s \cos\phi_s \\
  k_{iy} &= k_w \cos\theta_i \sin\phi_i & k_{sy} &= k_w \cos\theta_s \sin\phi_i\\
  k_{iz} &= -k_w \sin\theta_i & k_{sz} &= k_w \sin\theta_s \, .
\end{align}
Without loss of generality, the assumption $\phi_i=0$ can be made for isotropic roughness (which we only consider here). For 3D scattering geometry and 2D roughness, grazing angles $\theta_i,\theta_s$ have support $[0,\pi/2]$, whereas the azimuthal angle has support $\phi_s \in [0,2\pi]$. For 2D scattering geometry and 1D roughness, $\phi_s=\pi$  and $\theta_i,\theta_w \in [0,\pi]$, which renders the $y$ component of the wave vector zero.

The horizontal component of the Bragg wave vector, $\Delta \mathbf{K}$ and Bragg wavenumber magnitude $\Delta K$ in 2D are
\begin{align}
  \Delta \mathbf{K} &= \mathbf{K}_s - \mathbf{K}_i \\
  \Delta K &= \sqrt{(\Delta K_x)^2 + (\Delta K_y)^2}\, .
\end{align}
The Bragg wavenumber in 1D is
\begin{align}
  \Delta K_x = k_{sx} - k_{ix} \, .
\end{align}

\begin{figure}
  \centering
  \includegraphics[width=3.375in]{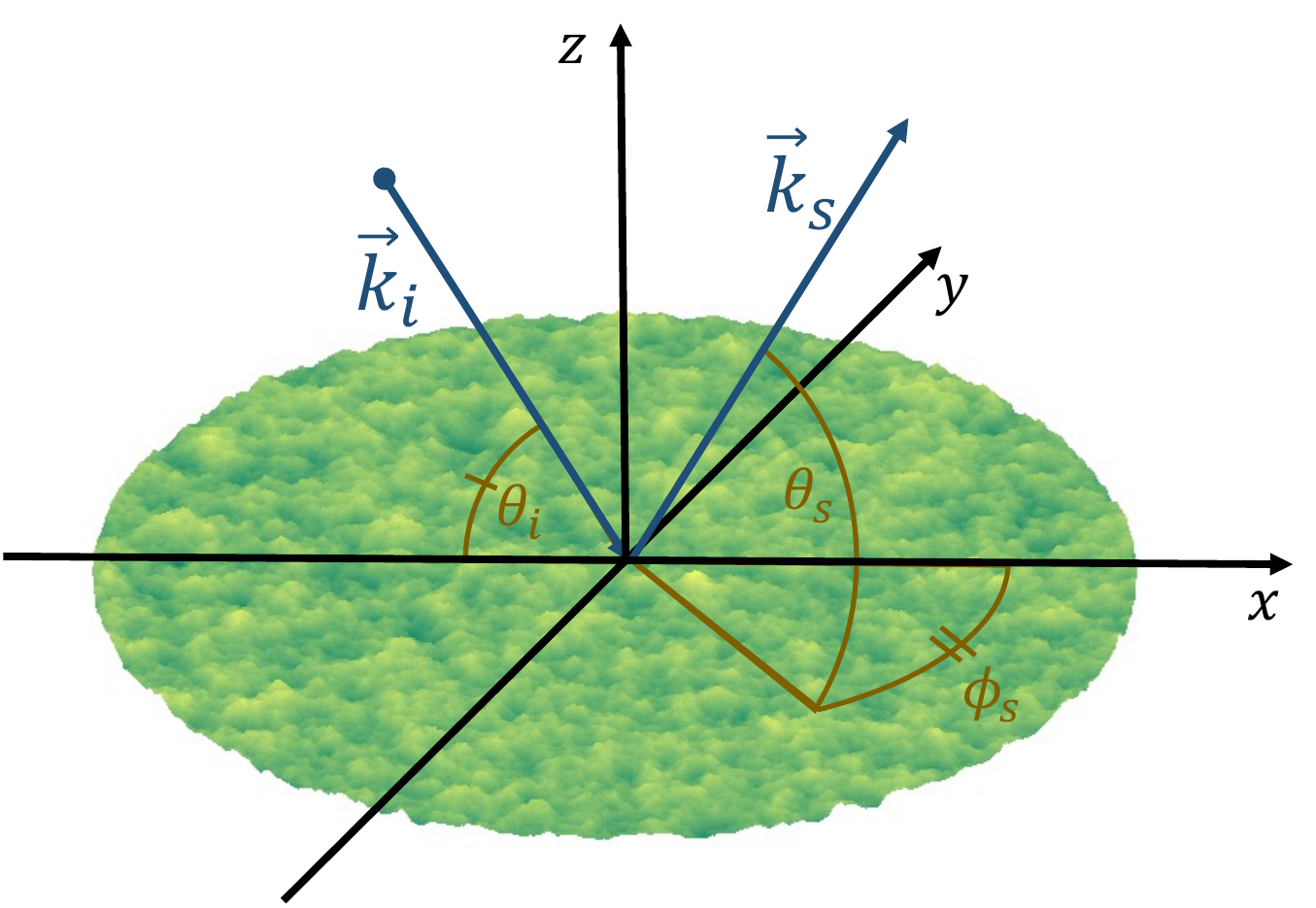}
  \label{fig:geometry}
  \caption{(color online) Geometry of the scattering problem under consideration for two-dimensional roughness and three-dimensional scattering geometry. The rough interface is shown in green and yellow. Incident and scattered wave vectors are shown in blue. The incident grazing angle,$\theta_i$, scattered grazing angle, $\theta_s$, and scattered azimuthal angle, $\phi_s$ are shown in brown, and a right-handed coordinate system is shown in black.}
\end{figure}

For 2D roughness, the interface is specified by $f(x,y)$, where $x,y$ are the two horizontal coordinates. For 1D roughness, we use $f(x)$, where $x$ is the single horizontal coordinate. For all cases, Gaussian height statistics and wide-sense stationary is assumed. A sample rough interface is shown in Fig.~\ref{fig:geometry}, along with the incident and scattered wave vectors, and their angles.

The Fourier transforms of each of these rough interfaces are specified by
\begin{align}
	  F(K_x) &= \frac{1}{2\pi}\int f(x)e^{\mathrm{i}K_x x}\,\mathrm{d}x \\
  F(\mathbf{K}) &= \frac{1}{(2\pi)^2}\int f(\mathbf{R})e^{\mathrm{i}\mathbf{K}\cdot \mathbf{R}}\,\mathrm{d}^2R \, ,
\end{align}
where $\mathbf{K}= (K_x,K_y)$ is the horizontal wavenumber vector of the roughness spectrum,  $\mathbf{R}=(x,y)$ is the 2D horizontal coordinate vector, and $F$ is the complex amplitude spectrum of the rough interface. $F$ is related to the power spectrum $W$ by
\begin{align}
	 W_1(K_x) &= \langle F(K_x)F^\ast(K_x^\prime)\rangle \delta(K_x - K_x^\prime) \\
  W_2(\mathbf{K}) &= \langle F(\mathbf{K})F^\ast(\mathbf{K}^\prime)\rangle \delta(\mathbf{K} - \mathbf{K}^\prime) \, ,
\end{align}
where $\delta(x)$ is the Dirac delta function.

The roughness covariance functions are denoted $C_1$ and $C_2$ for 1D and 2D roughness spectra respectively. They are defined by
\begin{align}
	  \langle f(x)f(x_0)\rangle &= C_1(x-x_0)\\
      \langle f(x,y)f(x_0, y_0)\rangle &= C_2(x-x_0,y-y_0) \, ,
\end{align}
and are related to their respective power spectra by
\begin{align}
W_2(\mathbf{K}) &= \frac{1}{(2\pi)^2}\int C_2(\mathbf{R})e^{\mathrm{i}\mathbf{K}\cdot \mathbf{R}}\,\mathrm{d}^2R \\
W_1(K_x) &= \frac{1}{2\pi}\int C_1(x)e^{\mathrm{i}K_x x}\,\mathrm{d}x\, .
\end{align}

For azimuthally isotropic roughness, the 2D power spectrum is related to the 2D covariance function by a Fourier-Hankel transform of order zero,
\begin{align}
  W_2(K_r) = \int_0^\infty R J_0(K_r R )C_2(R)\mathrm{d}R.
  \label{eq:hankelTransform}
\end{align}
where $R = \sqrt{x^2 + y^2}$ is the horizontal coordinate magnitude, and $K_r = \sqrt{K_x^2 + K_y^2}$ is the radial wavenumber magnitude. For 2D roughness, only isotropic forms of the roughness covariance and power spectrum are considered in this work.

\subsection{Roughness covariance models}
The Gaussian covariance function is commonly used to model statistical roughness with a single length scale, often for the purposes of model validation \cite{Thorsos1988,Thorsos1989,Broschat1997}. The covariance functions are given for 1D and 2D roughness by,
\begin{align}
  C_1(x) &= h_1^2 e^{-x^2/L_1^2} \label{eq:gauss_1d_cov}\\
  C_2(x,y) &= C_2(R) = h_2^2 e^{-R^2/L_2^2}\label{eq:gauss_2d_cov}
\end{align}
where $L_1$ is the correlation length for 1D roughness, and $L_2$ is the correlation length for 2D roughness. The notation $C_2(R) = C_2(x,y)$ implies an assumption of azimuthal isotropy. These forms have Fourier and Hankel transforms respectively
\begin{align}
  W_1(K_x) &= \frac{h_1^2 L_1 }{2\sqrt{\pi}} e^{-K_x^2 L_2^2/4} \label{eq:gauss_1d_spectrum}\\
  W_2(K_r) &= \frac{h_2^2 L_2^2}{4\pi} e^{-K_r^2 L_2^2/4}\, . \label{eq:gauss_2d_spectrum}
\end{align}

%\subsection{Exponential covariance}
The exponential covariance function is often a better model for natural roughness than the Gaussian form, since it is a power-law spectrum at high wavenumbers, with an exponent of -2 for 1D roughness and -3 for 2D roughness\cite{Jackson2007}. The von K{\'a}rm{\'a}n form given in Ch. 6 of \cite{Jackson2007} is more general than the exponential form (since it allows for a variable exponent), although the exponential form is a special case of it. The exponential covariance for 1D and 2D roughness are
\begin{align}
  C_1(x) &= h_1^2 e^{-|x|K_{01}} \label{eq:exp_1d_cov} \\
  C_2(R) &= h_2^2 e^{-R K_{02}}\,. \label{eq:exp_2d_cov}
\end{align}
Again, isotropy is assumed for the 2D case. These covariance functions have Fourier/Hankel transforms respectively
\begin{align}
  W_1(K_x) &= \frac{w_1}{K_{01}^2 + K_x^2} \label{eq:exp_1d_spectrum}\\
  W_2(K_r) &= \frac{w_2}{\left(K_{02}^2 + K_r^2 \right)^{3/2}}\, .\label{eq:exp_2d_spectrum}
\end{align}
where $w_2=h_2^2 K_{02}/(2\pi)$ is the 2D spectral strength, and $w_1 = h_1^2K_{01}/\pi$ is the 1D spectral strength.

\subsection{The Kirchhoff Integral}
The Kirchhoff integral (KI) is defined here for 1D and 2D roughness. Although there are different conventions for this integral, the form of the 2D version used here is from \cite{Jackson2020}, and the 1D version is from \cite{Olson2020a}. The KI for 2D isotropic roughness is,
\begin{align}
  I_K^{2D}(\eta) = e^{-\eta h_2^2}\int_0^\infty u J_0\left(\frac{\Delta K u}{k_w}\right) \left[ e^{\eta C_2(u/k_w)} - 1  \right] \, \mathrm{d}u
  \label{eq:kirchh_int_2d}
\end{align}
and The KI for 1D roughness is
\begin{align}
  I^{1D}_K(\eta) = 2 e^{-\eta h_1^2} \int_0^\infty \cos\left(\frac{\Delta K_x u}{k_w} \right) \left[e^{\eta C_1(u/k_w)} -1 \right] \,\textrm{d}u
  \label{eq:kirchh_int_1d}\, .
\end{align}

In the Kirchhoff approximation \cite{Ishimaru1978a}, and small slope approximations for impenetrable boundaries \cite{Voronovich1985}, and interfaces between fluid \cite{Voronovich1999}, elastic \cite{Yang1994}, and dielectric boundaries \cite{Voronovich_1994}, $\eta = \Delta k_z^2$ , where $\Delta k_z = k_{sz} - k_{iz}$. However, in the layered small slope approximation \cite{Jackson2020}, $\eta$ is the product between two different quantities, that are related to the vertical component of the wavenumber in each domain. Therefore, $\eta$ can be a complex scalar. In numerical tests, only the case of $\eta = \Delta k_z^2$ is considered, although the proposed method is valid for the general complex case.

It is illustrative to show the convergence rate of the Kirchhoff integral computed using the trapezoidal rule. For the trapezoidal rule, there are two dimensionless numbers that must be selected, $u_{max}$, the upper limit of the integral (since the semi-infinite domain must be truncated), and $\delta u$, the sampling interval. The number of sampling points is $N_u = u_{max}/\delta u$.
This number is used as a proxy for the computational cost of computing the trapezoidal rule. A true accounting of this cost would estimate the number of floating point operations, or flops, since evaluation of the integrand at each point may take many flops, especially evaluation of special functions, such as the Bessel function of the first kind. $u_{max}$ should be large enough such that the exponential functions in Eqs.~(\ref{eq:kirchh_int_2d}) and,(\ref{eq:kirchh_int_1d}) approach unity, so that they cancel with the second term in square brackets. This situation occurs when $u_{max} \gg k_w/K_{0n}$, where $K_{0n}$ represents either $K_{01}$ or $K_{02}$. The sampling parameter should be small enough that it captures oscillations in the cosine or Bessel function. For the KI convention used here, oscillations are smallest for the specular direction, for which $\Delta K$ or $\Delta K_x$ is zero, and increase away from the specular direction.

\begin{figure}
\centering
\includegraphics[width=3.375in]{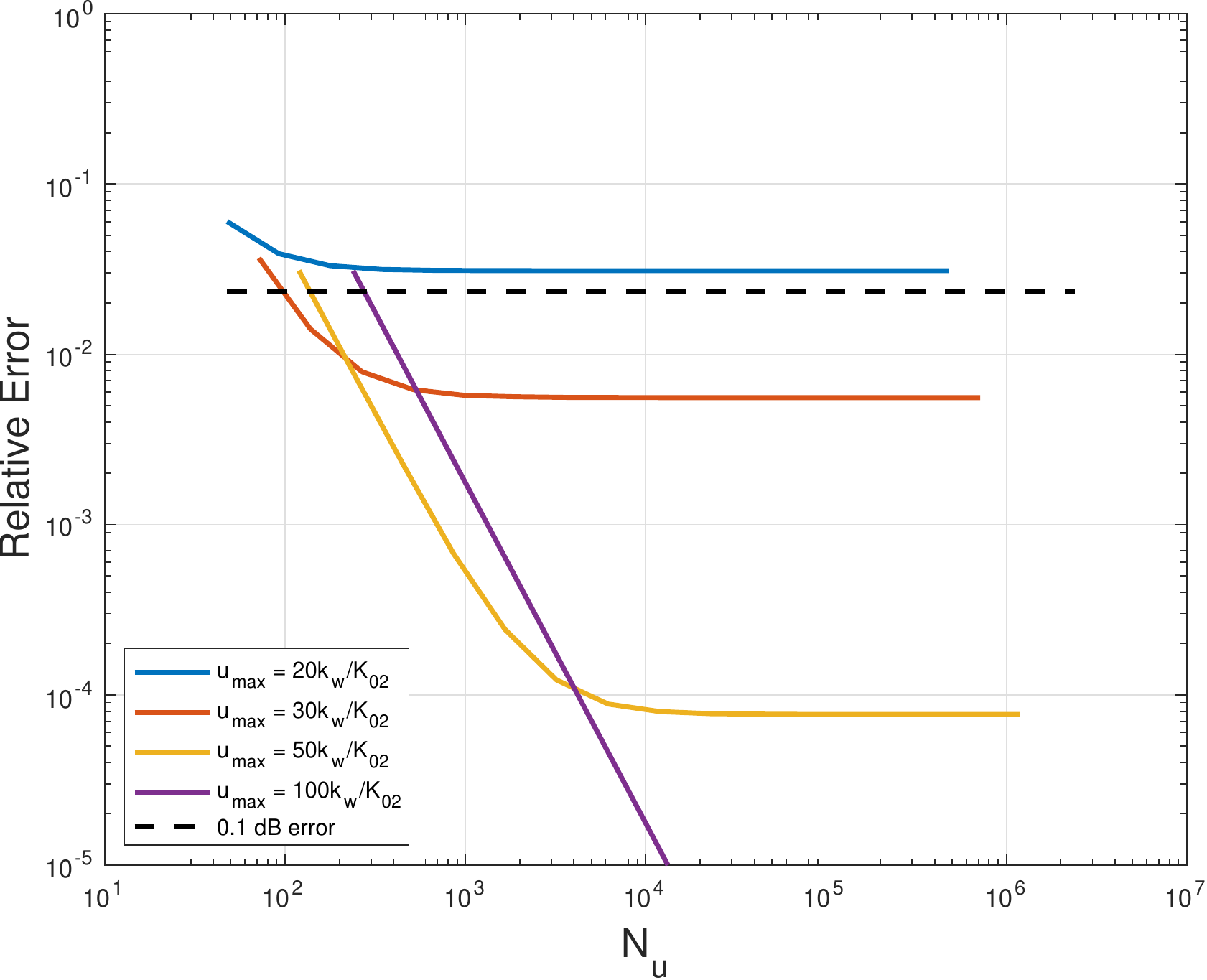}
\caption{(color online) Convergence rate expressed in relative error as a function of number of samples $N_u$ used for the trapezoidal rule, for several values of $u_{max}$ specified in the legend. The incident and scattered grazing angles were 80$^\circ$, the wavenumber was 8.37 rad/m. Rms height was 1/$k_w$, and $K_{02}/k_w=0.05$. The 2D Kirchhoff integral was used for this figure.}
\label{fig:kirchof_convergence}
\end{figure}

The convergence rate of as a function of $N_u$ for several values of $u_{max}$ is plotted in Fig.~\ref{fig:kirchof_convergence}. Four solid lines are used to plot the relative accuracy of the 2D Kirchhoff integral using exponential covariance, for $k_wh_2 = 1$, $K_{02}/k_w=0.05$, and $\theta_i=\theta_s=80^\circ$, and $\phi_s= \pi$. Specific values of $u_{max}$ are marked in the legend. The sampling interval $\delta u$ was varied, and the relative error is plotted as a function of $N_u$. A reference line of 0.1 dB  (2.33\%) relative error is plotted for reference. Over 100 points are required to obtain convergence within 2.33\%. If $h_2$, the incident and scattered grazing angles, or $K_{02}$ are decreased, the number of points required increases. For example, if $\theta_i$ and $\theta_s$ are both changed to $45^\circ$, then $u_{max}$ should be greater than 50$k_w/K_{02}$, and 1000 points are required to achieve 0.1 dB accuracy. It is thus advantageous to have an approximation of this integral. It is also interesting to note that if $u_{max}$ is not large enough, then decreasing $\delta u$ has little effect on the accuracy.

Given this difficulty in numerically evaluating the Kirchhoff integral, there have been previous efforts to approximate it. In previous approximations to the 2D Kirchhoff integral \cite{Dashen1990,Drumheller2001,Gragg2001}, the limit of $K_{02}\to0$ was taken. While this may be an adequate approximation for very high frequency systems, or environment with a very large outer scale, it may not be appropriate in every case. In this limit the rms height, $h_2$ ,tends to infinity. The coherent field scattered by the surface therefore has zero intensity, and all intensity is incoherent. For cases where there is some finite coherent intensity, this scenario is unacceptable, especially if energy conservation between the coherent and incoherent scattered field must be enforced.

To illustrate the need to model finite values of the cutoff parameter, the backscattering cross section for several values of the wavenumber cutoff is plotted. The spectral strength was set to $w_2=10^{-6}$ m. The sound speed ratio, $\nu$ was set to 1.17, and the density ratio $a_{\rho}$ was set to 2.2 with the complex sound speed ratio $a_c=\nu/(1+0.01i)$. The frequency was set to $20$ kHz, and the sound speed was $1500$m/s. The small-slope approximation for a fluid half-space was used to compute the scattering cross section, using formulae in Ch. 13 of \cite{Jackson2007}.

\begin{figure}
  \centering
  \includegraphics[width=3.375in]{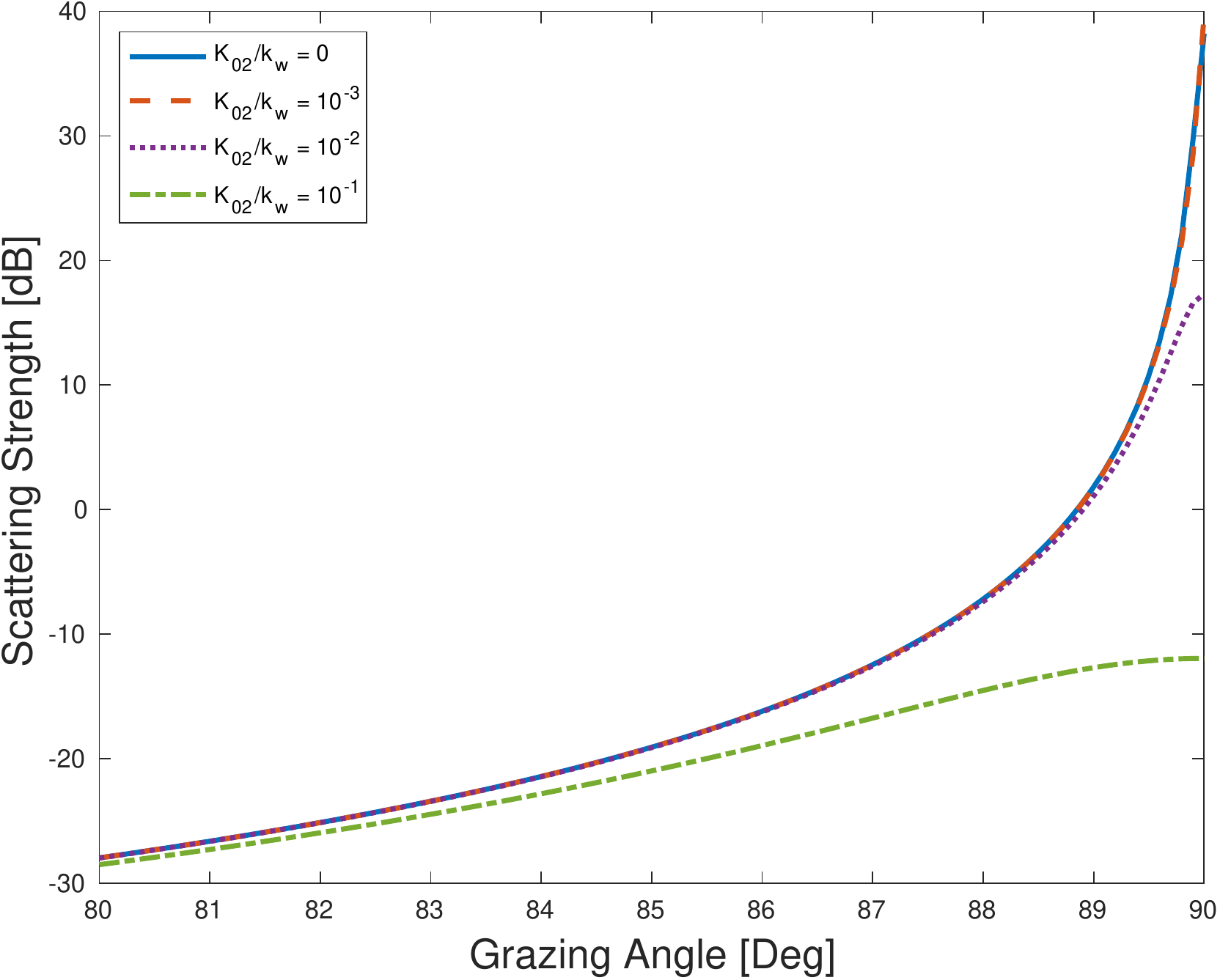}
  \caption{(color online) Scattering strength dependence on angle and spectral cutoff. Several values of the wavenumber cutoff for the 2D Kirchhoff integral were used, which are marked in the legend. Note the great disparity near normal incidence for these four cases.}
  \label{figs:ss_cutoff_example}
\end{figure}

The outer scale was set to specific values of 0, $k_w$/1000, $k_w$/100, and $k_w/10$, and the backscattering cross section near the specular direction is plotted in Fig.~\ref{figs:ss_cutoff_example}. The case of $K_{02}=0$ was computed using the approximations of \cite{Gragg2001}, and the finite $K_{02}$ cases were computed using the trapezoidal integration rule. At the specular direction, there is an enormous difference in scattering strength, from -12 dB to 39 dB.
From this figure, it can be concluded that only if the ratio $k_w /K_{02}$ is less than or equal to $10^{-3}$ does the limit of a pure power law produce accurate results. If the specular region is required for modeling or inversion of scattering strength (such as for multibeam systems \cite{Hellequin2003,Pouliquen1992,Hefner2015,Sternlicht2003a}), then a fast method is needed to compute the Kirchhoff integral for finite power-law cutoff parameters. Note that if $w_2$ is increased, then the differences between these four cases becomes smaller, as the specular peak is less sharp than it is for the value of $w_2$ in this figure.
%Such a method is developed in the next section.

\section{Functional Taylor series approximation}
\label{sec:series}
The form of the KI is that of a Fourier or Hankel transform, where the function to be transformed is the exponential of the roughness covariance times $\eta$. In both the 1D and 2D Kirchhoff integrals, the term in brackets does not have an analytic Fourier or Hankel transform. The fact that the argument of the exponential function has an analytic Fourier or Hankel transform (it is just the power spectrum) can be exploited to yield an efficient approximation to this integral. To achieve this approximation, a functional Taylor series of the integral is used, since the Kirchhoff integral can be viewed as a functional, which takes a function as an argument, and returns a real or complex number. The function argument is the roughness covariance in this case.

\subsection{Functional Taylor series definition}
To make these results applicable to both 1D and 2D scenarios, notation is introduced to make the functional aspect of the Kirchhoff integral explicit. Let $F_K[a(u)]$ be the functional at the core of the Kirchhoff integral.  The function $a(u)$ is the argument of the functional $F$, which is defined as
\begin{align}
  F_K[a(u)] = \int_0^\infty g(u)e^{a(u)}\textrm{d}u\, ,
\end{align}
where $a(u)$ is $\eta C_n(u/k_w)$, $C_n(u/k_w)$ is either the 1D or 2D isotropic covariance function, and $g(u)$ is either $\cos(\Delta k_x u /k_w)$ for the 1D integral, or $uJ_0(\Delta K u/k_w)$ for the 2D integral.

A functional analog of the Talor series can be constructed if the functional agrument is transformed to $a(u) + \epsilon b(u)$, with $\epsilon$ a positive real number. Then, an ordinary Taylor series of $F$ can be made in powers of $\epsilon$, leaving the parameter $\epsilon$ arbitrary for now \cite{Dreizler_1990}
\begin{align}
  F_K(a + \epsilon b) = \sum\limits_{m=1}^{\infty}\frac{1}{m!} \left[\frac{\textrm{d}^m F(a + \epsilon b)}{\mathrm{d}\epsilon^m}\right]_{\epsilon=0} \epsilon^m\, ,
  \label{eq:functional_taylor_series}
\end{align}
where it is implied that $a=a(u)$ and $b=b(u)$ are functions in some function space. The expansion is around a reference function $a(u)$ in terms of some function $b(u)$. It is instructive to compare the functional Taylor series to an ordinary one. In the Taylor series for functions, $a(u)$ plays the role of the point at which the derivatives are taken in an ordinary Taylor series, and $b(u)$ plays the role of the point at which the series should be computed. The parameter $\epsilon$ can be arbitrary, as it is the product $\epsilon b$ that must be small for the series to converge quickly.

Since the derivatives are with respect to a real number, $\epsilon$, the definition of $F_K[a(u)]$ can be used in the definition of the ordinary derivative to calculate these derivatives, resulting in
\begin{align}
  \frac{\mathrm{d}^m F(a + \epsilon b)}{\mathrm{d}\epsilon^m } &= \int_0^\infty g(u)\frac{\mathrm{d}^m}{\mathrm{d}\epsilon^m} e^{a + \epsilon b}\textrm{d}u \\
  &= \int_0^\infty g(u) b^m e^{a}e^{\epsilon b} \textrm{d}u \, .
\end{align}
Setting $\epsilon=0$ and substituting this back into the functional Taylor series,
\begin{align}
  F_K(a + \epsilon b) = \sum\limits_{m=0}^{\infty}\frac{\epsilon^m}{m!}
\int_0^\infty g(u)b^m e^{a} \textrm{d}u\, .
\end{align}
To have an approximate representation of the Kirchhoff integral, the functions $a(u)$ and $b(u)$, and real number $\epsilon$ must be specified. An analytically tractable solution is possible if $a(u)$ is chosen as the zero function (a function which returns zero for any input value $u$), $b(u) = \eta C_n(u/k_w)$, and $\epsilon =1$. Notice that these three choices makes the functional $F_K(\eta C(u/k_w)]$, which is the functional component of the Kirchhoff integral of interest here.

Making these substitutions into the 2D Kirchhoff integral results in
\begin{align}
  \begin{split}
 I^{2D}_K(\eta) &= e^{-\eta h_2^2}\sum\limits_{m=0}^{\infty} \frac{1}{m!} \int_0^\infty u J_0(\Delta K u/k_w) \\
 &\times \left[\eta^l C^m_2(u/k_w)\right] \mathrm{d}u\\
   &- e^{-\eta h_2^2}\int_0^\infty u J_0(\Delta K u/k_w)\mathrm{d}u \label{eq:2d_series_generic}
\end{split} \\
\begin{split}
 &= e^{-\eta h_2^2}\sum\limits_{m=1}^{\infty} \frac{\eta^m}{m!} \int_0^\infty u J_0(\Delta K u/k_w)\\
 &\times C^m_2(u/k_w)\mathrm{d}u \label{eq:2d_series_generic_simple}
\end{split}
\end{align}
where (\ref{eq:2d_series_generic_simple}) results from the fact that $C_2^0=1$, $\eta^0=1$, and $0!=1$, and cancels with the second additive term in (\ref{eq:2d_series_generic}). In 1D, the Kirchhoff integral becomes, after similar manipulations,
\begin{align}
  \begin{split}
  I^{1D}_K(\eta) &= 2e^{-\eta h_1^2}\sum\limits_{m=1}^{\infty} \frac{\eta^m}{m!} \\
  &\times \int_0^\infty \cos(\Delta k_x u /k_w) C_1^m(u/k_w)\mathrm{d} u
  \end{split}
\end{align}
To compute this series in practice, a partial sum with upper limit $M$ is used. Thus far, no specific covariance has been assumed.

This series results in integrals that involve integer powers of the roughness covariance multiplied by a vertical wavenumber parameter, $\eta$. Thus to take advantage of the functional Taylor series, integer powers of the roughness covariance function must have analytic Fourier/Hankel transforms. This property is not true for the most popular seafloor roughness model--the von K{\'a}rm{\'a}n spectrum \cite{Jackson2007}. However, the exponential covariance and Gaussian spectrum both have this property. While these two specific roughness models do not cover the wide range of spectral forms seen in natural roughness, they are useful for model validation (where a Gaussian \cite{Thorsos1988,Thorsos1989,Broschat1997} or exponential \cite{Olson2020a} models are commonly assumed), or for when the seafloor has an approximately exponential roughness covariance (i.e. a power law spectal exponent of 3 for 2D roughness).

A fruitful area of future work may be to use a numerical integral in the wavenumber domain for the general von K{\'a}rm{\'a}n spectrum, since the Fourier transform of the $n$-th power of a function in the spatial domain is an $n$-fold convolution in the wavenumber domain. This aspect is not explored here, although it shows promise, since the $n$-fold convolution is non-oscillatory and may be computationally cheap to compute, compared to the Kirchhoff integral directly. Another possibility is to approximate arbitrary covariance functions with weighted sums of exponential or Gaussian functions, which has been used elswhere in the scattering literature \cite{Winebrenner_1985,Hefner_2014}.

%The functional Taylor series that is used here is for the exponential function, namely \cite{Rytov}
%\begin{align}
%  e^{a f(x)} \approx \sum\limits_{\ell=0}^\infty \frac{a^\ell f^\ell(x)}{\ell!}
%\end{align}

%Plugging this expansion into Eq.~(\ref{eq:kirchh_int_2d}) yields,
%\begin{align}
%  \begin{split}
%  I_K^{2D}(\eta) &= e^{-\eta h_2^2}  \\
%  & \times \int\limits_0^\infty u J_0(u\Delta K/k_w) \left[\sum\limits_{l=1}^\infty \frac{\eta^\ell}{\ell!} C_2^\ell(u/k_w)-1\right] \,\textrm{d}u
%\end{split}
%\end{align}

%Since the first term in brackets is unity for $\ell=0$, this term cancels with the second term in brackets, and we are left with,
%\begin{align}
%  \begin{split}
%  I_K^{2D}(\eta) &= e^{-\eta h_2^2}  \\
%  & \times \int\limits_0^\infty u J_0(u\Delta K/k_w) \sum\limits_{l=1}^\infty \frac{\eta^\ell}{\ell!} C_2^\ell(u/k_w) \,\textrm{d}u
%\end{split}
%\end{align}
\subsection{Specific forms of the roughness covariance}
If the exponential form for the 2D roughness covariance is substituted into (\ref{eq:2d_series_generic_simple}), the following integral is obtained:
\begin{align}
  \begin{split}
  I_K^{2D}(\eta) &= e^{-\eta h_2^2} \sum\limits_{m=1}^M\frac{\eta^m }{m!}h_2^{2m} \\
   &\times \int\limits_0^\infty u J_0(u\Delta K/k_w) e^{-m u K_{02}/k_w}\,\textrm{d}u \, .
\end{split}
\end{align}

This integral can be performed analytically using  Eqs.~(\ref{eq:hankelTransform}),(\ref{eq:exp_2d_cov}), and (\ref{eq:exp_2d_spectrum}). A modified form of the covariance with the effective spectral cutoff, $K_{02}^\prime = m K_{02}$ is used to complete the integral, resulting in
\begin{align}
  I_K^{2D}(\eta) &= e^{-\eta h_2^2} \sum\limits_{m=1}^M \frac{\eta^m h_2^{2m}}{m!} \frac{k_w^2 K_{02} m}{\left[ (K_{02}m)^2 + \Delta K^2\right]^{3/2}}
  \label{eq:KI_series_exp_2d}
\end{align}
This series converges rapidly, owing to the factorial in the denominator. Even if $\eta h_2^2$ is large, the factorial term, $m!$ grows faster than $\eta^m h_x^{2m}$. Additionally, contributions of the second fraction are small when $m K_{02} \gg \Delta K$. A similar process is used to obtain an expression using the 1D exponential covariance.
\begin{align}
  I_K^{1D}(\eta) &= 2e^{-\eta h_1^2} \sum\limits_{m=1}^M \frac{\eta^m h_1^{2m} }{m!} \frac{k_w K_{01} m}{ (K_{01}m)^2 + \Delta K_x^2}
  \label{eq:KI_series_exp_1d}
\end{align}
Following similar steps, the 2D Kirchhoff integral under the Gaussian covariance can be computed using Eqs.~(\ref{eq:hankelTransform}),(\ref{eq:gauss_2d_cov}), and (\ref{eq:gauss_2d_spectrum}),
\begin{align}
  I_K^{2D}(\eta) &= e^{-\eta h_2^2} \sum\limits_{m=1}^M \frac{\eta^m h_2^{2m}}{m!} \frac{k_w^2 L_2^2}{2m}e^{-\Delta K^2 L_2^2/(4m)} \, .
  \label{eq:KI_series_gauss_2d}
\end{align}
The 1D version for Gaussian covariance is
\begin{align}
  I_K^{1D}(\eta) &= 2e^{-\eta h_1^2} \sum\limits_{m=1}^\infty \frac{\eta^m h_1^{2m} }{m!} \frac{k_w L_1\sqrt{\pi}}{2\sqrt{m}} e^{-\Delta k_x^2L_1^2/(4m)} \, .
  \label{eq:KI_series_gauss_1d}
\end{align}

\section{Convergence Tests and rate of convergence}
\label{sec:convergence}
\subsection{Asymptotic Convergence}
\label{subse:asymptoticConvergence}
The series approximations developed in Section~\ref{sec:series} can be demonstrated to converge unconditionally. This property is shown here using the ratio test. Let $a_m$ be the $m$th terms in any of Eq.~(\ref{eq:KI_series_exp_2d}), (\ref{eq:KI_series_exp_1d}), (\ref{eq:KI_series_gauss_2d}), or (\ref{eq:KI_series_gauss_1d}). If the ratio
\begin{align}
  r =\lim_{m\to\infty} \frac{a_{m+1}}{a_m}
\end{align}
is strictly less than 1, then the series converges \cite{Courant1987}. Forming this ratio for the 2D exponential covariance, we have
\begin{align}
  r = \lim_{m\to\infty} \frac{\eta h_2^2\left[(K_{02}m)^2 + \Delta K^2\right]^{3/2} }{m\left[(K_{02}(m+1))^2 + \Delta K^2\right]^{3/2} }
\end{align}
If $m$ is set to be much greater than $K_{02}/\Delta K$, then the ratio of the bracketed terms tends to unity. Therefore, the limit can be easily calculated as
\begin{align}
    r = \lim_{m\to\infty} \frac{\eta h_2^2}{m} =0\, .
\end{align}
The series approximation for the 2D Kirchhoff integral with exponential converges independently of any of the roughness or geometry parameters. Very similar steps can be used to prove that the 1D version converges unconditionally as well.

Turning to the 2D Gaussian case, the ratio is
\begin{align}
  r = \lim_{m\to\infty}\frac{\eta h_2^2m}{(m+1)^2}\exp\left[\frac{\Delta K^2L_2^2}{4}\left(\frac{-1}{m+1}+\frac{1}{m}\right) \right]\, .
\end{align}
The term $(-1/(m+1) + 1/m)$ tends to zero as $m$ tends to infinity, so the ratio can be written as
\begin{align}
  r = \lim_{m\to\infty}\frac{\eta h_2^2m}{(m+1)^2} = 0
\end{align}
which also converges unconditionally. Very similar steps can be used to prove the same result for the 1D Gaussian case.

The ratio $r$ is the convergence rate of the series. Computing the error by keeping $M$ terms requires an estimate of the remainder terms, which can be difficult. A simpler way to truncate the series is to use $r$ to specify a minimum change by keeping one additional term in the series. Then, once $r$ is set (e.g., to 0.01 or whatever the user requires), the value of $m$ that satisfies that requriement can be found. Then $M$ is set to be this value of $m$.
%The convergence properties shown above result from the Taylor series involving the exponential function, and the properties of the covariance functions used. The coefficients of the series approximation for the exponential with $F_K[a)u]$ decay very quickly. Each term in the resulting approximate KI series is a form of the power spectral density with a modfied cutoff or correlation length. Since the Gaussian and exponential functions both have spectra that decay with wavenumber, the coefficients \cdots.

\subsection{Numerical examples of convergence}
\label{sec:numericalConvergence}
It was demonstrated in the previous subsection that these series converge unconditionally. However, the rate of convergence is also important, which governs how many terms should be kept to achieve a specified error. If hundreds or thousands of terms are required, then the series may not be faster to compute than a direct numerical integral. Additionally, the factorial suffers from numerical overflow at $m=171$ for double precision floating point numbers, so it is not practical to compute partial sums with very large $M$ (at least for a straightforward implementation of this series).

Two examples of the numerical accuracy of the series methods are given, one for the exponential covariance, and another for the Gaussian covariance. Both examples are one-dimensional, although similar convergence rates were observed for the 2D case. Both covariance forms have two parameters, $h_1$ and $K_0$ for the exponential model, and $h_1$ and $L_1$ for the Gaussian model. Non-dimensional values for both of these parameters are set using the acoustic wavenumber in water, $k_w$. The specific values for $k_w h_1$ are $\left[0.1, 1, 2, 5 \right]$. The specific values values for the length (or inverse length) scales, $K_0/k_w$, are $\left[0.05, 0.1, 1 \right]$. For the Gaussian covariance, we use the same values for length scale as the exponential covariance, $k_w L_1 = k_w/K_0$. The series method is computed using a maximum value for $m$, $M=120$.

\begin{figure*}
  \includegraphics[width=0.8\textwidth]{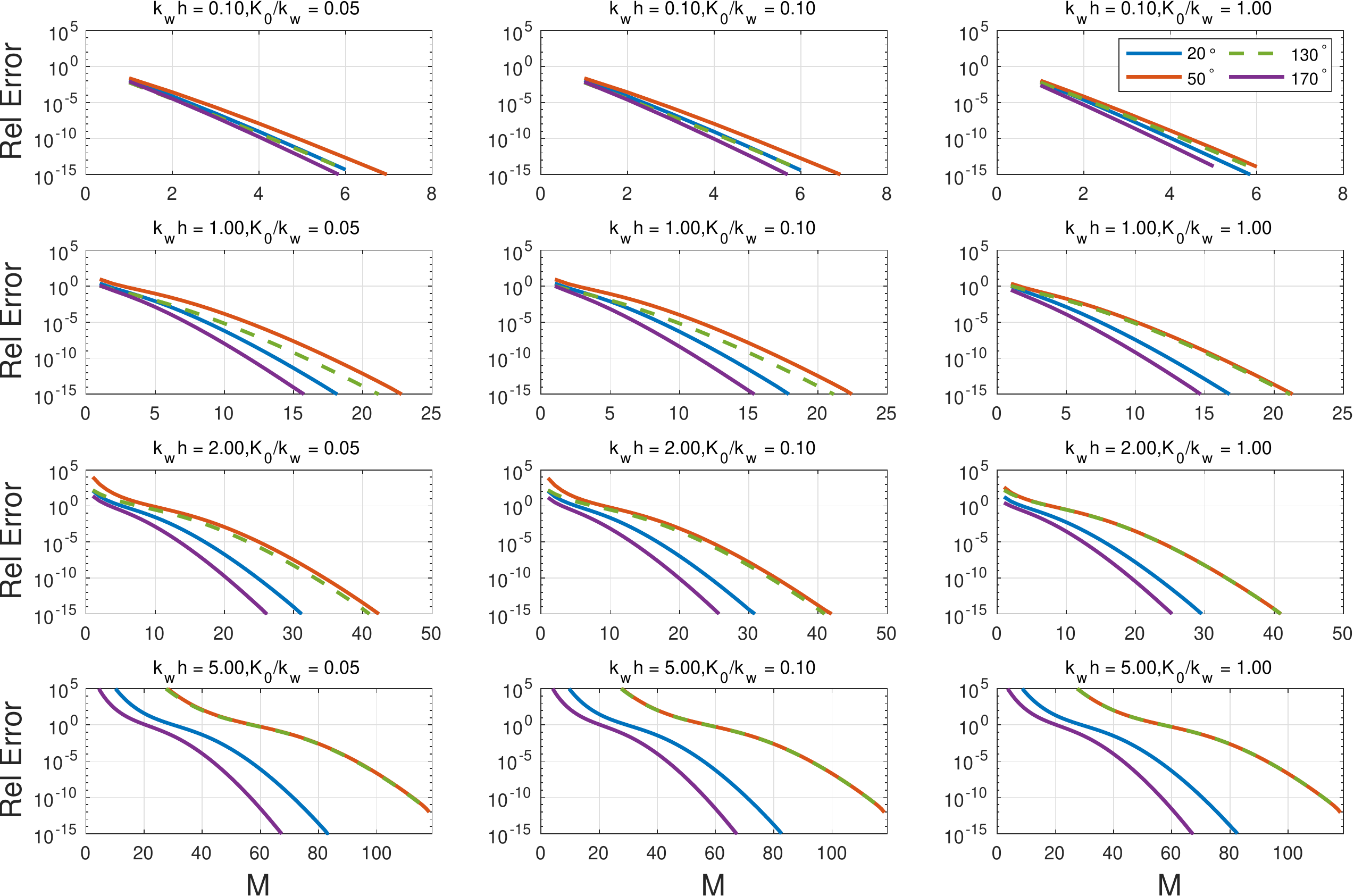}
  \caption{(color online) Illustration of the convergence rate for the series approximation to the Kirchhoff integral for 1D exponential covariance, for several values of $\theta_s$, as a function of $k_w h_1$, and $K_0/k_w$. Each roughness parameter is plotted in its own subfigure. The horizontal axes denotes $M$, and the vertical axis is relative error. Note that the ordinate is different for each set of $kh$ values.}
  \label{fig:exp1d_convergence_rate}
\end{figure*}

\begin{figure*}
  \includegraphics[width=0.8\textwidth]{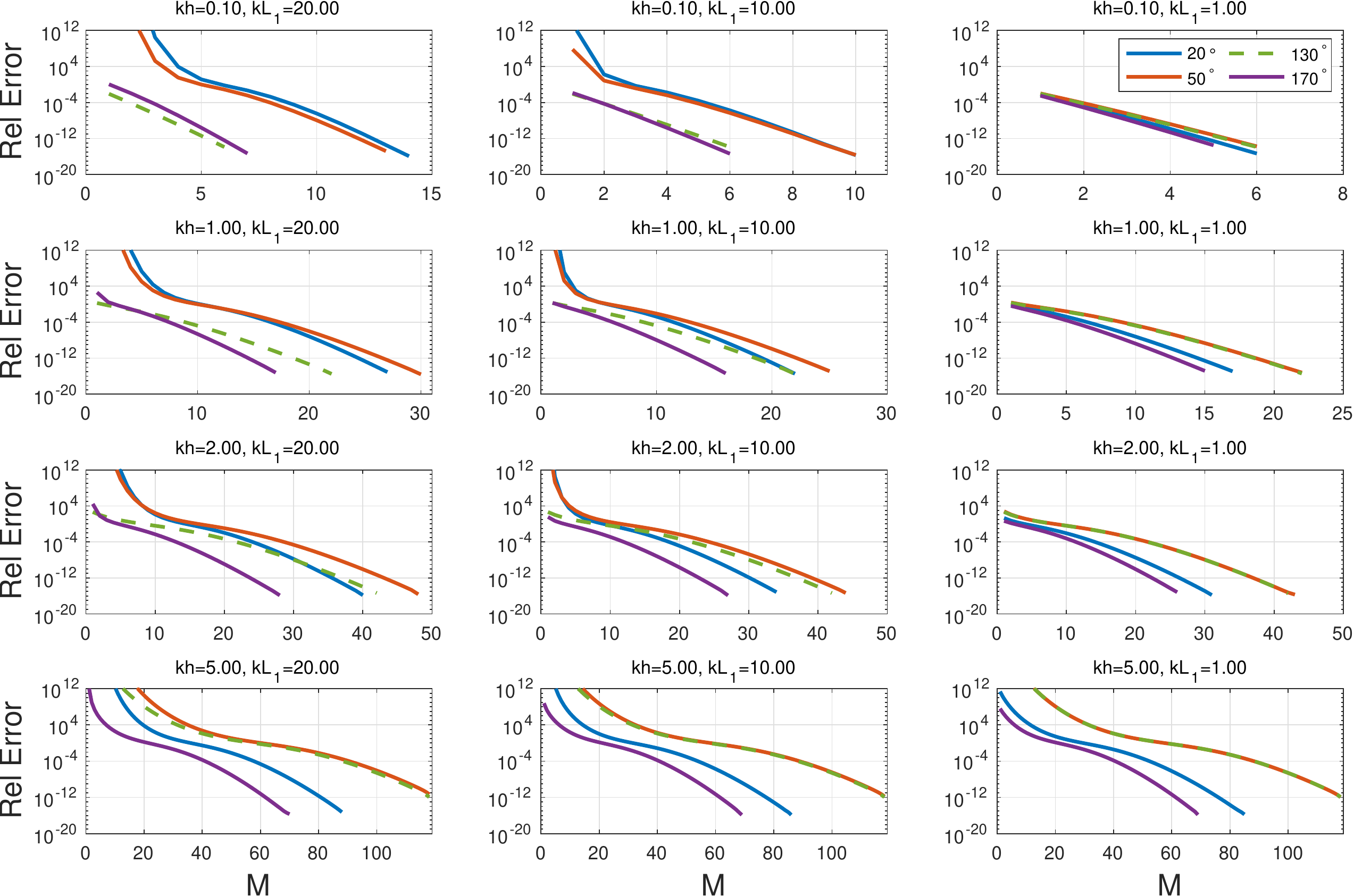}
  \caption{(color online) Illustration of the convergence rate for the series approximation to the Kirchhoff integarl for 1D Gaussian covariance for several values of $\theta_s$, as a function of $k_w h_1$, and $K_0/k_w$. Each roughness parameter is plotted in its own subfigure. The horizontal axes denotes $M$, and the vertical axis is relative error. Note that the ordinate is different for each set of $kh$ values.}
  \label{fig:gauss1d_convergence_rate}
\end{figure*}

The benchmark used here for the Kirchhoff integral is the trapezoidal rule, using an upper limit of $u_{max}= 50 k_w /K_{01}=50 k_w L_1$. Therefore,  50 correlation lengths / outer scales are captured. The $u$ axis has a sampling interval $\delta u$ of $10^{-4}$, which is about 1500 points per wavelength. For the smallest values of $K_{01}$, these parameters result in $10^{7}$ points for the trapezoidal integral rule.
For the plots of the partial series (Figs.~\ref{fig:exp1d_convergence_rate} and \ref{fig:gauss1d_convergence_rate}),
$\theta_i = 50^\circ$, and $\theta_s = \left[20^\circ,50^\circ,130^\circ,170^\circ\right]$. The frequency was set to 2 kHz, but any value of frequency could be used, as the roughness parameters are set in a non-dimensional fashion. One notable difference between the 1D and 2D case was that the cosine term has constant magnitude as a function of $u$, whereas the $uJ_0(u)$ term in the 2D integral grows as $u$ tends to infinity. Smaller $u_{max}$ was required for the 1D case than the 2D case.

%First, results for the exponential covariance are presented. A comparison of the numerical integration method, series method, and the absolute value of the difference between the two is presented in Fig.~\ref{fig:exp1d_error}. In all cases, the series with $M=120$ produces a very accurate answer, compared with the numerical integral. In the cases with small $k_w h$, the absolute error is limited by the numerical method, rather than the series truncation. In the left column, top two plots, the jagged error plot (dotted line) indicates that the numerical integration is limited by machine precision. Increasing the maximum value of $u$, or decreasing $\delta u$ resulted in the same absolute error.

% \begin{figure*}
%   \includegraphics[width=0.8\textwidth]{figs/exp1d_error.eps}
%   \caption{Comparison of the numerical integral, series approximation, and the difference between the two. Each roughness parameter is plotted in its own subfigure. The exponential covariance for the 1D Kirchhoff integral is shown.}
%   \label{fig:exp1d_error}
% \end{figure*}

%Show rate of convergence for three grazing angles, and plot $M$ on the horizontal axis. Do this for a bunch of parameters. It will be a better illustration.

The exponential case is presented first. A plot of the relative error of the series approximation is presented in Fig.~\ref{fig:exp1d_convergence_rate}. Each subfigure contains a different combination of $k_wh_1$, and $K_{01}/k_w$, with increasing $k_wh_1$ going from top to bottom, and increasing $K_{02}$ going from left to right. The parameter $M$ is varied between 1 and 120, and is the abcissa of these plots. The relative error between the the current $M$ and the numerical solution is plotted on the ordinate. Four scattered grazing angles are shown here, and are plotted as different lines denoted in the legend. Here, $50^\circ$ is the backscattering direction, $130^\circ$ is the specular direction, and the other two angles represent low back and forward scattering.

The dependence of the scattered grazing angle shows the dependence on $\Delta K_x$ and $\eta = \Delta k_z^2$. Convergence is fast for the smallest $k_w h_1 = 0.01$. Very high accuracy is achieved by only four terms. After 7 terms, the series converges to within double precision. As $k_w h_1$ is increased, the convergence rate is slower, and more terms are required to achieve acceptable accuracy. Using $M=120$, relative accuracy better than $10^{-5}$ is achieved for all roughness parameters and angles studied here.

%RThe results are similar, although the error is more influenced by numerical precision. In the upper left figure, the blue solid and red dashed lines overlap very well for scattered grazing angles abot 75$^\circ$, below this angle, agreement is much larger, with the series apparantly underestimating the integral. However, for those angles, the numerical method is limited by numerical precision. Decreasing the sampling interval or increasing the upper limit does not change the value of the Kirchhoff integral. It would seem that the oscillatory nature of the integral prevents accurate evaluation, when the Kirchhoff integral has a small magnitude.

% \begin{figure*}
%   \includegraphics[width=0.8\textwidth]{figs/gauss1d_error.eps}
%   \caption{Comparison of the numerical integral, series approximation, and the difference between the two. Each roughness parameter is plotted in its own subfigure. The Gaussian covariance for the 1D Kirchhoff integral is shown.}
%   \label{fig:gauss1d_error}
% \end{figure*}

The convergence rate for 1D Gaussian covariance is shown in Fig.~\ref{fig:gauss1d_convergence_rate}. In this figure, the error computed using the series solution with $M= 170$ as the reference integral, instead of the trapezoidal rule, due to the problems with numerical roundoff. The trapezoidal rule in some cases suffered from numerical roundoff error, since the absolute value of the Kirchhoff integral was far below double precision limit. This case occurred for very small rms height, and angles far away from the specular direction. The behavior of the series approximation is similar to the exponential covariance case, although the case of small $k_w h_1$ and large $k_w L_1$ in the upper left corner converges much more slowly for backscattering angles. This slow convergence is due to the exponential term in Eq.(\ref{eq:KI_series_gauss_1d}), which grows monotonically with increasing $m$. It has an upper limit of unity, but may grow quickly for small $m$ if $L_1$ is large. The role of the effective power spectrum is much more pronounced here. Note that this figure has different limits on the ordinate.

\section{Conclusion}
\label{sec:conclusion}
For two simple roughness covariance models, a functional Taylor series approximation to the Kirchhoff integral is possible in both 1D and 2D. This series converges quickly when the rms height is small compared to the wavelength, and slowly when it is big, but converges asymptotically for all roughness parameters. For practical purposes, the series converges to a high degree of accuracy using 120 or fewer terms in the partial sum. This approximation is much faster than direct numerical integration of the Kirchhoff integral, which is both oscillatory and slowly converging. It is limited to classes of roughness convariance functions, $C(x)$, where $C^n(x)$, has an analytical Fourier transform and $n$ is an integer  (\textit{mutatis mutandis} for the 2D case), which limits the applicability of this technique. Future work on this problem for other roughness covariance types would broaden the applicability of this technique.

% \bibliographystyle{jasaauthyear2}
% \bibliography{Jabref.bib}

\begin{thebibliography}{31}
\def\enquote#1{``#1,''}
\def\plainquote#1{``#1''}
\expandafter\ifx\csname natexlab\endcsname\relax\def\natexlab#1{#1}\fi
\providecommand{\dourl}[1]{\href{http://#1}{\nolinkurl{#1}}}
\providecommand{\bibinfo}[2]{#2}
\providecommand{\noopsort}[1]{}
\providecommand{\switchargs}[2]{#2#1}
  \def\eatspace #1{#1}

\bibitem[{Afifi \emph{et~al.}(2014)Afifi, Dusseaux, and Berrouk}]{Afifi2014}
\bibinfo{author}{Afifi, S.}, \bibinfo{author}{Dusseaux, R.},  and
  \bibinfo{author}{Berrouk, A.} (\textbf{\bibinfo{year}{2014}}).
  \enquote{\bibinfo{title}{Electromagnetic scattering from 3d layered
  structures with randomly rough interfaces: Analysis with the small
  perturbation method and the small slope approximation}}
  \bibinfo{journal}{{IEEE} Transactions on Antennas and Propagation}
  \textbf{62}(10), \bibinfo{pages}{5200--5208},
  \dodoi{10.1109/tap.2014.2341704}.

\bibitem[{Ainslie(2010)}]{Ainslie2010}
\bibinfo{author}{Ainslie, M.} (\textbf{\bibinfo{year}{2010}}).
  \emph{\bibinfo{title}{Principles of Sonar Performance Modelling}}
  (\bibinfo{publisher}{Springer Berlin Heidelberg}).

\bibitem[{Beckmann and Spizzichino(1987)}]{Beckmann1987}
\bibinfo{author}{Beckmann, P.},  and \bibinfo{author}{Spizzichino, A.}
  (\textbf{\bibinfo{year}{1987}}). \emph{\bibinfo{title}{The Scattering of
  Electromagnetic Waves from Rough Surfaces}} (\bibinfo{publisher}{Artech House
  Radar Library}).

\bibitem[{Berrouk \emph{et~al.}(2014)Berrouk, Duss{\'e}aux, and
  Afifi}]{Berrouk2014}
\bibinfo{author}{Berrouk, A.}, \bibinfo{author}{Duss{\'e}aux, R.},  and
  \bibinfo{author}{Afifi, S.} (\textbf{\bibinfo{year}{2014}}).
  \enquote{\bibinfo{title}{Electromagnetic wave scattering from rough layered
  interfaces: Analysis with the small perturbation method and the small slope
  approximation}} \bibinfo{journal}{Progress In Electromagnetics Research B}
  \textbf{57}, \bibinfo{pages}{177--190}, \dodoi{10.2528/pierb13101802}.

\bibitem[{Broschat and Thorsos(1997)}]{Broschat1997}
\bibinfo{author}{Broschat, S.~L.},  and \bibinfo{author}{Thorsos, E.~I.}
  (\textbf{\bibinfo{year}{1997}}). \enquote{\bibinfo{title}{An investigation of
  the small slope approximation for scattering from rough surfaces. {Part II}.
  numerical studies}} \bibinfo{journal}{J. Acoust. Soc. Am.} \textbf{101},
  \bibinfo{pages}{2615 -- 2625}.

\bibitem[{Courant(1987)}]{Courant1987}
\bibinfo{author}{Courant, R.} (\textbf{\bibinfo{year}{1987}}).
  \emph{\bibinfo{title}{Differential and Integral Calculus}},
  \bibinfo{volume}{\textbf{I}}, \bibinfo{edition}{2} ed.
  (\bibinfo{publisher}{Blackie}).

\bibitem[{Dashen \emph{et~al.}(1990)Dashen, Henyey, and Wurmser}]{Dashen1990}
\bibinfo{author}{Dashen, R.}, \bibinfo{author}{Henyey, F.~S.},  and
  \bibinfo{author}{Wurmser, D.} (\textbf{\bibinfo{year}{1990}}).
  \enquote{\bibinfo{title}{Calculations of acoustic scattering from the ocean
  surface}} \bibinfo{journal}{The Journal of the Acoustical Society of America}
  \textbf{88}(1), \bibinfo{pages}{310--323},
  \dodoi{http://dx.doi.org/10.1121/1.399953}.

\bibitem[{Dreizler and Gross(1990)}]{Dreizler_1990}
\bibinfo{author}{Dreizler, R.~M.},  and \bibinfo{author}{Gross, E. K.~U.}
  (\textbf{\bibinfo{year}{1990}}). \emph{\bibinfo{title}{Density Functional
  Theory}} (\bibinfo{publisher}{Springer Berlin Heidelberg}).

\bibitem[{Drumheller and Gragg(2001)}]{Drumheller2001}
\bibinfo{author}{Drumheller, D.~M.},  and \bibinfo{author}{Gragg, R.~F.}
  (\textbf{\bibinfo{year}{2001}}). \enquote{\bibinfo{title}{Evaluation of a
  fundamental integral in rough-surface scattering theory}}
  \bibinfo{journal}{J. Acoust. Soc. Am.} \textbf{110}(5),
  \bibinfo{pages}{2270--2275}, \dodoi{10.1121/1.1412445}.

\bibitem[{Gragg and Wurmser()}]{Gragg2005}
\bibinfo{author}{Gragg, R.~F.},  and \bibinfo{author}{Wurmser, D.} ().
  \enquote{\bibinfo{title}{Scattering from a rough seafloor with
  stratification}} in \emph{\bibinfo{booktitle}{Proceedings 2005 Conference on
  Boundary Influences in High Frequency, Shallow Water Acoustics}},
  \bibinfo{publisher}{Institute of Physics}.

\bibitem[{Gragg \emph{et~al.}(2001)Gragg, Wurmser, and Gauss}]{Gragg2001}
\bibinfo{author}{Gragg, R.~F.}, \bibinfo{author}{Wurmser, D.},  and
  \bibinfo{author}{Gauss, R.~C.} (\textbf{\bibinfo{year}{2001}}).
  \enquote{\bibinfo{title}{Small-slope scattering from rough elastic ocean
  floors: General theory and computational algorithm}} \bibinfo{journal}{The
  Journal of the Acoustical Society of America} \textbf{110}(6),
  \bibinfo{pages}{2878--2901}, \dodoi{10.1121/1.1412444}.

\bibitem[{Hefner(2015)}]{Hefner2015}
\bibinfo{author}{Hefner, B.~T.} (\textbf{\bibinfo{year}{2015}}).
  \enquote{\bibinfo{title}{Inversion of high frequency acoustic data for
  sediment properties needed for the detection and classification of uxos}}
  \bibinfo{type}{Final Report},
  \dourl{https://www.serdp-estcp.org/content/download/34593/333838/file/MR-2229-FR.pdf}.

\bibitem[{Hefner and Jackson(2014)}]{Hefner_2014}
\bibinfo{author}{Hefner, B.~T.},  and \bibinfo{author}{Jackson, D.~R.}
  (\textbf{\bibinfo{year}{2014}}). \enquote{\bibinfo{title}{Attenuation of
  sound in sand sediments due to porosity fluctuations}} \bibinfo{journal}{The
  Journal of the Acoustical Society of America} \textbf{136}(2),
  \bibinfo{pages}{583--595}, \dodoi{10.1121/1.4889864}.

\bibitem[{Hellequin \emph{et~al.}(2003)Hellequin, Boucher, and
  Lurton}]{Hellequin2003}
\bibinfo{author}{Hellequin, L.}, \bibinfo{author}{Boucher, J.~M.},  and
  \bibinfo{author}{Lurton, X.} (\textbf{\bibinfo{year}{2003}}).
  \enquote{\bibinfo{title}{Processing of high-frequency multibeam echo sounder
  data for seafloor characterization}} \bibinfo{journal}{IEEE J. Ocean Eng.}
  \textbf{28}(1), \bibinfo{pages}{78--89}, \dodoi{10.1109/JOE.2002.808205}.

\bibitem[{Ishimaru(1978)}]{Ishimaru1978a}
\bibinfo{author}{Ishimaru, A.} (\textbf{\bibinfo{year}{1978}}).
  \emph{\bibinfo{title}{Wave Propagation and Scattering in Random Media}},
  \bibinfo{volume}{\textbf{I}} (\bibinfo{publisher}{Academic Press}).

\bibitem[{Jackson and Olson(2020)}]{Jackson2020}
\bibinfo{author}{Jackson, D.},  and \bibinfo{author}{Olson, D.~R.}
  (\textbf{\bibinfo{year}{2020}}). \enquote{\bibinfo{title}{The small-slope
  approximation for layered, fluid seafloors}} \bibinfo{journal}{The Journal of
  the Acoustical Society of America} \textbf{147}(1), \bibinfo{pages}{56--73},
  \dourl{https://doi.org/10.1121/10.0000470}, \dodoi{10.1121/10.0000470}.

\bibitem[{Jackson and Richardson(2007)}]{Jackson2007}
\bibinfo{author}{Jackson, D.~R.},  and \bibinfo{author}{Richardson, M.~D.}
  (\textbf{\bibinfo{year}{2007}}). \emph{\bibinfo{title}{High-Frequency
  Seafloor Acoustics}} (\bibinfo{publisher}{Springer}, \bibinfo{address}{New
  York, NY}).

\bibitem[{Lurton(2010)}]{Lurton2010}
\bibinfo{author}{Lurton, X.} (\textbf{\bibinfo{year}{2010}}).
  \emph{\bibinfo{title}{An Introduction to Underwater Acoustics: Principles and
  Applications}}, \bibinfo{edition}{2nd} ed. (\bibinfo{publisher}{Springer},
  \bibinfo{address}{Berlin}).

\bibitem[{Olson and Jackson(2020)}]{Olson2020a}
\bibinfo{author}{Olson, D.~R.},  and \bibinfo{author}{Jackson, D.}
  (\textbf{\bibinfo{year}{2020}}). \enquote{\bibinfo{title}{Scattering from
  layered seafloors: Comparisons between theory and integral equations}}
  \bibinfo{journal}{The Journal of the Acoustical Society of America}
  \textbf{148}(4), \bibinfo{pages}{2086--2095}, \dodoi{10.1121/10.0002164}.

\bibitem[{Pouliquen(1992)}]{Pouliquen1992}
\bibinfo{author}{Pouliquen, E.} (\textbf{\bibinfo{year}{1992}}).
  \enquote{\bibinfo{title}{Identification des fonds marins superficiels a
  l'aide de signaux d'echo-sounders}} Ph.D. thesis, \bibinfo{school}{University
  of Paris VII}.

\bibitem[{Steininger \emph{et~al.}(2013)Steininger, Holland, Dosso, and
  Dettmer}]{Steininger2013}
\bibinfo{author}{Steininger, G.}, \bibinfo{author}{Holland, C.~W.},
  \bibinfo{author}{Dosso, S.~E.},  and \bibinfo{author}{Dettmer, J.}
  (\textbf{\bibinfo{year}{2013}}). \enquote{\bibinfo{title}{Seabed roughness
  parameters from joint backscatter and reflection inversion at the malta
  plateau}} \bibinfo{journal}{The Journal of the Acoustical Society of America}
  \textbf{134}(3), \bibinfo{pages}{1833--1842}, \dodoi{10.1121/1.4817833}.

\bibitem[{Sternlicht and de~Moustier(2003)}]{Sternlicht2003a}
\bibinfo{author}{Sternlicht, D.~D.},  and \bibinfo{author}{de~Moustier, C.~P.}
  (\textbf{\bibinfo{year}{2003}}). \enquote{\bibinfo{title}{Remote sensing of
  sediment characteristics by optimized echo-envelope matching}}
  \bibinfo{journal}{The Journal of the Acoustical Society of America}
  \textbf{114}(5), \bibinfo{pages}{2727--2743}, \dodoi{10.1121/1.1608019}.

\bibitem[{Thorsos(1988)}]{Thorsos1988}
\bibinfo{author}{Thorsos, E.~I.} (\textbf{\bibinfo{year}{1988}}).
  \enquote{\bibinfo{title}{The validity of the {K}irchhoff approximation for
  rough surface scattering using a {G}aussian roughness spectrum}}
  \bibinfo{journal}{J. Acoust. Soc. Am.} \textbf{83}, \bibinfo{pages}{78--92}.

\bibitem[{Thorsos and Broschat(1995)}]{Thorsos1995}
\bibinfo{author}{Thorsos, E.~I.},  and \bibinfo{author}{Broschat, S.~L.}
  (\textbf{\bibinfo{year}{1995}}). \enquote{\bibinfo{title}{An investigation of
  the small slope approximation for scattering from rough surfaces. {Part I}.
  theory}} \bibinfo{journal}{J. Acoust. Soc. Am.} \textbf{97},
  \bibinfo{pages}{2082 -- 2093}.

\bibitem[{Thorsos and Jackson(1989)}]{Thorsos1989}
\bibinfo{author}{Thorsos, E.~I.},  and \bibinfo{author}{Jackson, D.~R.}
  (\textbf{\bibinfo{year}{1989}}). \enquote{\bibinfo{title}{The validity of the
  perturbation approximation for rough surface scattering using a {Gaussian}
  roughness spectrum}} \bibinfo{journal}{J. Acoust. Soc. Am.} \textbf{86},
  \bibinfo{pages}{261--277}.

\bibitem[{Voronovich(1994)}]{Voronovich_1994}
\bibinfo{author}{Voronovich, A.} (\textbf{\bibinfo{year}{1994}}).
  \enquote{\bibinfo{title}{Small-slope approximation for electromagnetic wave
  scattering at a rough interface of two dielectric half-spaces}}
  \bibinfo{journal}{Waves in Random Media} \textbf{4}(3),
  \bibinfo{pages}{337--367}, \dodoi{10.1088/0959-7174/4/3/008}.

\bibitem[{Voronovich(1985)}]{Voronovich1985}
\bibinfo{author}{Voronovich, A.~G.} (\textbf{\bibinfo{year}{1985}}).
  \enquote{\bibinfo{title}{Small-slope approximation in wave scattering by
  rough surfaces}} \bibinfo{journal}{Sov. Phys. JETP} \textbf{62},
  \bibinfo{pages}{65--70}.

\bibitem[{Voronovich(1999)}]{Voronovich1999}
\bibinfo{author}{Voronovich, A.~G.} (\textbf{\bibinfo{year}{1999}}).
  \emph{\bibinfo{title}{Wave Scattering from Rough Surfaces}},
  \bibinfo{edition}{second} ed. (\bibinfo{publisher}{Springer},
  \bibinfo{address}{Berlin Heidelberg}).

\bibitem[{Voronovich(2002)}]{Voronovich_2002}
\bibinfo{author}{Voronovich, A.~G.} (\textbf{\bibinfo{year}{2002}}).
  \enquote{\bibinfo{title}{The effect of the modulation of bragg scattering in
  small-slope approximation}} \bibinfo{journal}{Waves in Random Media}
  \textbf{12}(3), \bibinfo{pages}{341--349},
  \dodoi{10.1088/0959-7174/12/3/306}.

\bibitem[{Winebrenner and Ishimaru(1985)}]{Winebrenner_1985}
\bibinfo{author}{Winebrenner, D.~P.},  and \bibinfo{author}{Ishimaru, A.}
  (\textbf{\bibinfo{year}{1985}}). \enquote{\bibinfo{title}{Application of the
  phase-perturbation technique to randomly rough surfaces}}
  \bibinfo{journal}{Journal of the Optical Society of America A}
  \textbf{2}(12), \bibinfo{pages}{2285}, \dodoi{10.1364/josaa.2.002285}.

\bibitem[{Yang and Broschat(1994)}]{Yang1994}
\bibinfo{author}{Yang, T.},  and \bibinfo{author}{Broschat, S.~L.}
  (\textbf{\bibinfo{year}{1994}}). \enquote{\bibinfo{title}{Acoustic scattering
  from a fluid{\textendash}elastic-solid interface using the small slope
  approximation}} \bibinfo{journal}{The Journal of the Acoustical Society of
  America} \textbf{96}(3), \bibinfo{pages}{1796--1804},
  \dodoi{10.1121/1.410258}.

\end{thebibliography}

\end{document}